\documentstyle[aps,prl]{revtex}
\begin{document}
\draft
\twocolumn[\hsize\textwidth\columnwidth\hsize\csname @twocolumnfalse\endcsname
\title{Spontaneous spin stripe dimerization in the doped  t-J model.}
\author{O.P. Sushkov\cite{a}}
\address{Max Planck Institute for Physics of Complex Systems, Nothnitzer
Str. 38, D-001187 Dresden, Germany}

\maketitle

\begin{abstract}
To investigate  spin dimerization in the $t-J$ model we 
consider the extended $t-J-\delta$ model with explicit spin dimerization 
introduced 
via parameter $\delta$. At zero doping the dimerized spin liquid is
unstable at $\delta \le \delta_c \approx 0.3$. We demonstrate that the
doping stabilizes the dimerized stripe phase: the $\delta_c(x)$ decreases 
when doping $x$ increases. At doping larger than critical,
$x > x_c$, the dimerized phase is stable even without explicit dimerization,
i.e. at $\delta=0$.

\end{abstract}

]

It is widely believed that the 2D $t-J$ model is relevant to the
low energy physics of high-temperature superconductors.
This is why investigation of this model is of great interest
both for theory and experiment. In spite of great efforts
during more than a decade
there is no full understanding of the phase diagram of the $t-J$ model,
however some facts are well established. At zero doping the model is equivalent
to the Heisenberg model on a square lattice which has long range 
Neel order \cite{O1}. Doping by holes destroys the order.
A simplified picture of noninteracting holes leads to the Neel state
instability with respect to spirals at arbitrary small but finite 
doping \cite{SS}.
However more sophisticated numerical calculations which take into
account renormalization of the hole Green's function under the doping
indicate that the Neel order is stable below some critical hole
concentration $x_c$ \cite{Ig}.
In the Neel phase ($x < x_c$), in all waves except s-wave, 
there is magnon mediated superconducting pairing
interaction between holes.
\cite{Fl}.
It is also clear that at very small hopping 
there is phase separation in the model because separation leads to reduction of
the number of destroyed antiferromagnetic links \cite{Emery}

The purpose of the present work is to elucidate spin structure of the 
ground state in quantum disordered regime, i.e at $x > x_c$. The most 
important hint comes from experiment:
indications of stripes in the high-$T_{c}$ materials \cite{stripes}.
Another important hint is a remarkable stability of the spin dimerized phase
in the frustrated $J_1-J_2$ model. The idea of such state for this model
was first formulated by Read and Sachdev \cite{Read}, and was then confirmed by
further work \cite{Gel,Kot}. The stability of
 such a configuration implies that the lattice symmetry is
 spontaneously broken and the ground state is four-fold
 degenerate.
Such a route towards quantum disorder is known rigorously
 to take place in one dimension, where the Lieb-Schultz-Mattis
(LSM) theorem guarantees that a gapped phase always breaks the
 translational symmetry.
Some time ago Affleck suggested that the
 LSM theorem  can be extended to higher dimensions, and
 the gapped states of quantum systems necessarily break the discrete 
 symmetries of the lattice \cite{Affleck}.
The example of the $J_1-J_2$ model provides further support for this idea.

There have been several attempts to consider the spin-dimerized phase in a 
doped
Heisenberg antiferromagnet. For this purpose Affleck and Marston \cite{Mar} 
analyzed Hubbard-Heisenberg model in the weak-coupling regime,
Grilli, Castellani and G. Kotliar \cite{Grilly} considered $SU(N)$, 
$N\to \infty$, $t-J$ model, and very recently
Vojta and Sachdev \cite{Voj} considered
$Sp(2N)$, $N\to \infty$,  $t-J$ model
with long range Coulomb interaction. These works indicated a stability of
the spin-dimerized phase in some region of parameters, providing a very
important guiding line. However relevance of these results to
 "physical regime" of the $t-J$ model remained unclear.
In the present work we demonstrate that the spin-dimer order is stable,
 the only small parameter used in the analysis is hole concentration with 
respect to the half filling.

The Hamiltonian under consideration is
 \begin{equation}
 \label{H}
 H=-t\sum_{\langle ij \rangle \sigma} c_{i\sigma}^{\dag}c_{j\sigma}
 + \sum_{\langle ij \rangle} J_{ij} \left({\bf S}_i{\bf S}_j-
 {1\over 4}n_in_j\right).
 \end{equation}
 $c_{i \sigma}^{\dag}$ is the  creation operator of an electron with
 spin $\sigma$ $(\sigma =\uparrow, \downarrow)$ at site $i$
 of the two-dimensional square lattice. The $c_{i \sigma}^{\dag}$ operators
 act in the Hilbert space with no double electron occupancy.
 The $\langle ij \rangle$ represents nearest neighbour sites.
The spin operator is ${\bf S}_i={1\over 2}\sum_{\alpha,\beta}
 c_{i \alpha}^{\dag} {\bf \sigma}_{\alpha \beta} c_{i \beta}$,
 and the number density operator is 
 $n_i=\sum_{\sigma}c_{i \sigma}^{\dag}c_{i \sigma}$. Antiferromagnetic
interactions $J_{ij}> 0$ are arranged in a stripe pattern shown in Fig. 1:
solid links correspond to $J_{ij}=J_{\perp}=J(1+\delta)$, and
dashed links correspond to $J_{ij}=j=J(1-\delta)$.
At half filling ($\langle n_i\rangle =1$) the Hamiltonian (\ref{H}) 
has already been studied\cite{Gel,Kat}: for
 $\delta > \delta_c \approx 0.303$ the ground state is a quantum disordered
state with gapped spectrum, and for  $\delta < \delta_c$ there is
spontaneous Neel ordering with gapless spin waves.

In order to study the stability of the dimer phase we first derive an effective
Hamiltonian in terms of bosonic operators creating spin-wave triplets (magnons)
$t^{\dag}_{i\alpha}$, $\alpha=x,y,z$ and fermionic operators creating holes 
$a^{\dag}_{\sigma}$, $\sigma =\uparrow, \downarrow$
from the spin singlets 
shown in Fig. 1.   
This Hamiltonian consists of four parts: the spin-wave part $H_t$,
the hole part $H_h$, the spin-wave-hole interaction $H_{th}$, and 
the hole-hole interaction $H_{hh}$. Let us start from $H_t$.
Similar effective theories have been
  derived in Refs.\cite{boson} and we only present the result:
  $ H_t = H_{2} + H_{3} + H_{4} + H_{U}$, where
$H_{2}  = \sum_{\bf{k}, \alpha} \left\{ A_{\bf{k}}
t_{\bf{k}\alpha}^{\dagger}t_{\bf{k}\alpha} +
\frac{B_{\bf{k}}}{2}\left(t_{\bf{k}\alpha}^{\dagger}
t_{\bf{-k}\alpha}^{\dagger} + \mbox{h.c.}\right) \right \}$,
$H_{3} =  \sum_{1+2=3} \mbox{R}({\bf k_{1}},{\bf k_{2}})
\epsilon_{\alpha\beta\gamma} t_{\bf{k_{1}}\alpha}^{\dagger}
t_{\bf{k_{2}}\beta}^{\dagger} t_{\bf{k_{3}}\gamma} + \mbox{h.c.}$, 
and
$H_{4}  =  \sum_{1+2=3+4} \mbox{T}({\bf k_{1}}-{\bf k_{3}})
 (\delta_{\alpha\delta}\delta_{\beta\gamma}-
\delta_{\alpha\beta}\delta_{\gamma\delta})
 t_{\bf{k_{1}}\alpha}^{\dagger}
t_{\bf{k_{2}}\beta}^{\dagger}t_{\bf{k_{3}}\gamma}
t_{\bf{k_{4}}\delta}$.
 We also introduce an infinite repulsion on each site, in order to
 enforce the kinematic constraint 
$t_{i \alpha}^{\dag} t_{i \beta}^{\dag} = 0$.
\begin{equation}
\label{hU} 
H_{U} = U \sum_{i,\alpha \beta} t_{i \alpha}^{\dagger}t_{i \beta}^{\dagger}
t_{i \beta}t_{i \alpha}, \ \ U \rightarrow \infty
\end{equation}
The following definitions are used in $H_2$:
$B_{\bf{k}}  = j(\cos{k_{y}}-0.5\cos{k_{x}})$ and
$A_{\bf{k}} =  J_{\perp} + B_{\bf k}$.
The matrix elements in the quartic and cubic interaction terms are:
$\mbox{T}({\bf k}) =  j(0.25 \cos{k_{x}} +0.5 \cos{k_{y}})$ and
$\mbox{R}({\bf p},{\bf q})=   0.25j(\sin{q_{x}} -\sin{p_{x}})$.
Throughout the paper we work in the Brillouin zone of the dimerized
 lattice.
 
At zero doping ($\langle n_i\rangle =1$)  $H_t$ is
an exact mapping of the original Hamiltonian (\ref{H}).
To analyse this case it is enough to apply the technique \cite{us,Kot}.
The result for the normal spin-wave Green's function reads:
\begin{equation}
\label{Gf}
G_{N}({\bf{k}},\omega) = {{\omega + \tilde{A}_{\bf{k}}(-\omega)}\over
{ \{ \omega + \tilde{A}_{\bf{k}}(-\omega) \} \{ \omega -
\tilde{A}_{\bf{k}}(\omega) \} +\tilde B_{\bf{k}}^{2}}}
\end{equation}
where $\tilde{A}_{\bf{k}}(\omega)= A_{\bf{k}} +\Sigma^N_4({\bf k})
+\Sigma_{Br}^{(1)}({\bf{k}},\omega)$ and
$\tilde{B}_{\bf{k}}(\omega)= B_{\bf{k}} +\Sigma^A_4({\bf k})$.
Normal $\Sigma_4^N$ and anomalous $\Sigma_4^A$ self-energies  are 
caused by the quartic 
interaction $H_4$ and the most important contribution
 $\Sigma_{Br}^{(1)}$ comes from the Brueckner diagrams as described in 
\cite{us}.
Strictly speaking there is also some contribution to the self-energy
caused by the "triple" interaction $H_3$. However this contribution
is very small (see, e.g. Ref.\cite{Kot}) and therefore we neglect it.

Expansion of the self-energy in powers of $\omega$ near $\omega=0$
gives quasiparticle residue and  spin-wave spectrum
$Z_{\bf{k}}^{-1} = 1 - \frac{\partial \Sigma_{Br}^{(1)}}{\partial\omega}$,
$\omega_{\bf{k}} = Z_{\bf{k}} \sqrt{[ \tilde{A_{\bf{k}}}(0)]^{2}
- \tilde{B_{\bf{k}}}^{2}}$.
Expressions for effective Bogoliubov parameters $u_{\bf k}$ and
$v_{\bf k}$ are given in \cite{us}.
The spin-wave gap $\Delta=\omega_{\bf k_0}$, ${\bf k_0}=(0,\pi)$,
obtained as a result of a selfconsistent
solution of Dyson's equations is plotted in Fig.2 (line at $x=0$). 
The critical value
of the explicit dimerization (point where the gap vanishes)
$\delta_c=0.298$ is in agreement with results of series
expansions \cite{Gel} and quantum Monte Carlo simulations \cite{Kat}.
The validity of the Brueckner approximation is justified by the
smallness of the gas parameter 
$n_t=\sum_{\alpha}\langle t^{\dag}_{i\alpha}t_{i\alpha}\rangle$.
At the critical point 
$n_t=0.13$.

Consider now doping by holes. On the single dimer $|s\rangle$ the hole can 
exist in symmetric and antisymmetric states. 
Because of hopping between the dimers there is mixing 
between these states, but the mixing is very small  (few per cent) and can be 
neglected.
The symmetric state has substantially lower energy and therefore only this 
state is
populated at doping. The corresponding hole creation operator $a^{\dag}$ 
is defined in the following way:
$\sqrt{2}a^{\dag}_{\sigma}|s\rangle=
(c^{\dag}_{2,\sigma}+ c^{\dag}_{1,\sigma})|0\rangle$,
where $1$ and $2$ numerate the dimer sites. Bare hole dispersion can be
found by calculating hopping matrix elements. This gives
$H_h=\sum(\epsilon_{\bf p}+const)a^{\dag}_{{\bf p}\sigma}a_{{\bf p}\sigma}$,
where the $const$ is chosen in such a way that
\begin{equation}
\label{ee}
\epsilon_{\bf p}=t(\cos p_y +0.5\cos p_x +1.5)
\end{equation}
vanishes at the minimum:  $\epsilon_{\bf p_0}=0$, ${\bf p_0}=(\pi,\pi)$.

The spin-wave-hole interaction $H_{th}$ can be easily calculated in the way
similar to that for doped spin-ladder \cite{Eder,sus}.
This interaction consists of two parts. The first one is interaction of a
hole and a magnon positioned at different dimers. This is a relatively weak
interaction which can be neglected \cite{com1}. The second part, which gives
the main effect, comes from the
constraint that a hole and a magnon can not coexist at
the same dimer: $t^{\dag}_{i\alpha}a^{\dag}_{i\sigma}=0$. 
To deal with this constraint we
introduce, similarly to (\ref{hU}), an infinite repulsion
\begin{equation}
\label{hU1} 
H_{U1} = U \sum_{i,\alpha \sigma} t_{i\alpha}^{\dagger}t_{i\alpha}
a_{i\sigma}^{\dagger}a_{i\sigma}
, \ \ U \rightarrow \infty.
\end{equation}
The exact hole-magnon scattering amplitude caused by this
interaction can be found via Bethe-Salpeter
equation shown in Fig.3a. It is similar to that for magnon-magnon 
scattering \cite{us}. The 
result is
\begin{equation}
\label{gam}
\Gamma(E,{\bf k})=-\left(\sum_{\bf q} {{Z_{\bf q}u_{\bf q}^2}\over
{E-\omega_{\bf q}-\epsilon_{\bf k-q}}}\right)^{-1},
\end{equation}
where $E$ and ${\bf k}$ is total energy and total momentum of the incoming
particles.

Let us denote the hole concentration by $x=n/N$, where $n$ is  number of holes,
and $N$ is number of sites. Hence on-site electron occupation number is
$\langle n_i\rangle=1-x$.
Concentration of holes in  terms of the dimerized lattice is two times larger
$n/(0.5N)=2x$, and this is the gas parameter of the
magnon-hole Brueckner approximation.
According to (\ref{ee}) the holes are concentrated in the
pocket in the vicinity of ${\bf p_0}=(\pi,\pi)$. Therefore the magnon
normal self-energy described by the diagram Fig. 3b is
\begin{equation}
\label{sh}
\Sigma_{Br}^{(2)}({\bf k},\omega)=2x \Gamma(\omega,{\bf k+p_0})
\end{equation}

It is instructive to consider first the case which allows an 
analytical solution: $J_{\perp} \gg j$, $\sqrt{2}\pi x \ll 1$. 
Bare magnon dispersion in this case is 
$\omega_{\bf k}\approx J_{\perp}+j(\cos k_y-0.5\cos k_x)$ and hence the 
integrals in (\ref{gam},\ref{sh}) can be calculated analytically with 
logarithmic accuracy. This gives
\begin{equation}
\label{spt}
\Sigma_{Br}^{(2)}({\bf k},\omega)\approx {{2\sqrt{2}\pi x (t+j)}\over
{\ln(12.5/\mu) +i\pi \theta(\delta\omega)}},
\end{equation}
where $\delta\omega = \left[\omega-\omega_{\bf k}+
j(\omega_{\bf k}-\omega_{\bf k_0})/(t+j)\right]\left. \right/(t+j)$,
 $\mu= \max(|\delta\omega|,\sqrt{2}\pi x)$, and
$\theta(\delta\omega)$ is a step function.
The magnon Green's function is
$G({\bf k},\omega)=
\left(\omega-\omega_{\bf k}-\Sigma_{Br}^{(2)}({\bf k},\omega)\right)^{-1}$.
For illustration the spectral function $Im G(\omega)$
at  ${\bf k}={\bf k_0}$, $t/j=3$ and different $x$ is plotted in Fig.4.
There are several conclusions from formula (\ref{spt}) and Fig. 4:
1) doping pushes the spin-wave spectrum up, 2) the effect is increasing with
hopping $t$, 3) finite width appears, 4) there is only 
a logarithmic dependence on the infrared cutoff.
Let us stress the importance of the point (4). It means that the
effect is practically independent of the long-range dynamics.
 Moreover, near the critical point ($\Delta=0$) the
 situation is even better: the spin-wave spectrum is linear
 and even the logarithmic divergence disappears.
Thus in the 2D case there is separation of scales which
justifies Brueckner approximation. If we 
tried to apply the described approach to the 1D case (say a doped spin ladder)
we would get into trouble: power infrared divergence appears in Brueckner 
diagram and  hence there is no justification for gas approximation.
Let us also comment on the point (3) (width). There is a
"triple" contribution  to the magnon self-energy, Fig. 3c.
This is a long-range contribution which is much less important than the 
Brueckner one, and this is why we neglect it (cf. with Ref.\cite{Kot}). 
However this diagram influences the width of the magnon spectral
function.

In the general case there are two contributions to the Brueckner self-energy:
$\Sigma_{Br}^{(1)}$, which is due to the magnon-magnon constraint, and
$\Sigma_{Br}^{(2)}$ which is due to the magnon-hole constraint.
To find the spin wave spectrum one has to solve selfconsistently
Dyson's equation for Green's function (\ref{Gf}), as it is described in Ref.
\cite{us}.  Results for the spin-wave "gap" $\Delta$  as a function of 
explicit dimerization $\delta$ for different hole concentrations $x$ and 
$t/J=3$ are plotted in Fig. 2. Strictly speaking
at $x \ne 0$ the $\Delta$ is not a gap because of the large decay width.
What we plot is the position of the centre of gravity of the magnon spectral
function.  However at  $\Delta \to 0$ the width vanishes, and therefore the 
critical regime is uniquely defined. When $\Delta$ is not small there is an 
interesting question of calculation of the exact shape of magnon spectral 
function. However to resolve this problem one needs to include 
long-range dynamics (diagram Fig.3c).

It is clear from Fig.2 that at $t/J=3$ and $x > x_c \approx 0.090$ the "gap" 
remains finite even at $\delta =0$.
This is regime of spontaneous dimerization.
For $t/J=2$ the critical concentration is $x_c=0.106$, and for $t/J=1$,
$x_c=0.132$. 
 Thus the doping stabilizes the dimerized 
phase. 
The larger the hopping $t$, the stronger the effect of stabilization.
(The same follows from  eq. (\ref{spt}).)
This statement is true only if $t/J \lesssim 10$. At $t/J \sim 10$
there is a crossover to quasiparticles with higher spin (hole-magnon 
bound states) \cite{sus} which indicate transition to
the Nagaoka regime.
The small parameter of the Brueckner approximation is concentration of 
holes in the dimerized lattice: $2x$. Therefore at
$t/J=3$ one should expect $\sim 20\%$  accuracy in calculation of $x_c$.
Note that the value of $x_c$ is close to that found in \cite{Ig} from the
Neel state.

An important effect related to the stabilisation
is suppression of spin wave quantum fluctuations by 
doping. At $x=0$ and $\delta=\delta_c=0.298$ the density
of spin fluctuations is $n_t\approx0.13$. 
Increasing of the hole concentration to $x=0.1$ (at $t/J=3$ and
$\delta=\delta_c$) gives $n_t \approx 0.02$.
At the critical point $t/J=3$, $\delta=0$, $x=x_c=0.09$ 
the density is just $n_t\approx 0.07$ (see also \cite{com1}).

The phase diagram of the $t-J-\delta$ model at
zero temperature is presented in Fig.5
Because of the mobile holes the dimerized spin liquid is a conducting state.
Stability of this state is a very robust
effect because it is due to the high energy correlations
(typical energy scale $\sim 2t$). There are also low energy (long range)
effects with typical energy scale $\sim 2tx$ which can lead
to hole-hole pairing, small amplitude density waves etc.
We do not consider these effects in the present work because they are 
secondary with respect to the main one: spin dimerization. 
It is also worth noting  that one can introduce additional
parameters $t'$ and $t''$ (next and next-next neighbour hopping) which
influence single hole dispersion (\ref{ee}),
shift position of the minimum and change shape of the "Fermi
surface":  stabilization is not sensitive to these details.

Strictly speaking one can not exclude  a possibility that at 
decreasing  $\delta$, at some point there is a first order phase transition 
to an absolutely different phase. However such a transition at zero
temperature always has a 
precursor: some specific bound state. The examples are: hole-magnon bound 
state for transition to Nagaoka regime in the present model at very large $t$,
multi-magnon bound states for a frustrated spin-ladder 
as a precursor of spinon deconfinement \cite{KSE}, and multi-magnon bound
states in 2D $J_1-J_2$-model at the 1st order transition point 
$J_2/J_1\approx 0.6$ \cite{Kot}. In the present model we do not see any hint
of such precursors and therefore we conclude that the first order phase
transition is highly unlikely.

We have discussed in detail the transition to the Neel state at small
 hole concentration $x$.
It is also clear that at large $x$ there is a 2nd order transition to normal 
Fermi liquid shown schematically in Fig. 5 by the dashed line. One can 
say that it happens when hole concentration in the dimerized lattice is 
equal to unity, $2x\sim 1$. However this is only a crude estimate.
Unfortunately we can not describe this part of the diagram more precisely
because our approach  assumes that $2x \ll 1$.

It has been shown recently that Heisenberg antiferromagnet 
on triangular lattice is very close to the instability with
respect to spontaneous dimerization \cite{Zheng}.
We would like to note that it is highly likely that in this model doping 
also stabilises the dimerized phase.

In conclusion, using the dilute gas approximation we have analyzed the 
phase diagram
of the $t-J-\delta$ model and the stability of spin-dimerized phase.
At doping larger than the critical one,
$x > x_c \approx 0.09$, the dimerized phase is stable even without 
explicit dimerization.

I thank V. N. Kotov and M. Yu. Kuchiev for stimulating discussions.

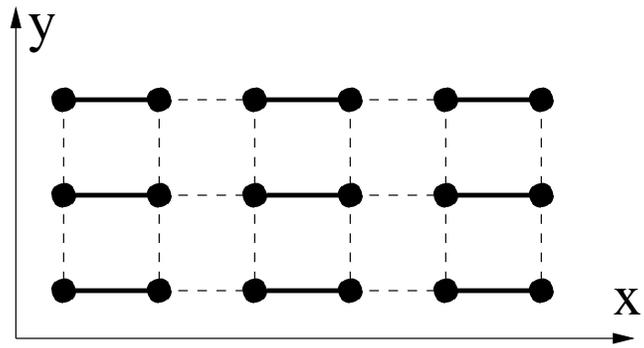
\begin{figure}
\caption
{Stripe spin dimerization on square lattice. Solid links
correspond to $J_{\perp}=J(1+\delta)$, and
dashed links correspond to $j=J(1-\delta)$.}
\label{fig.1}
\end{figure}

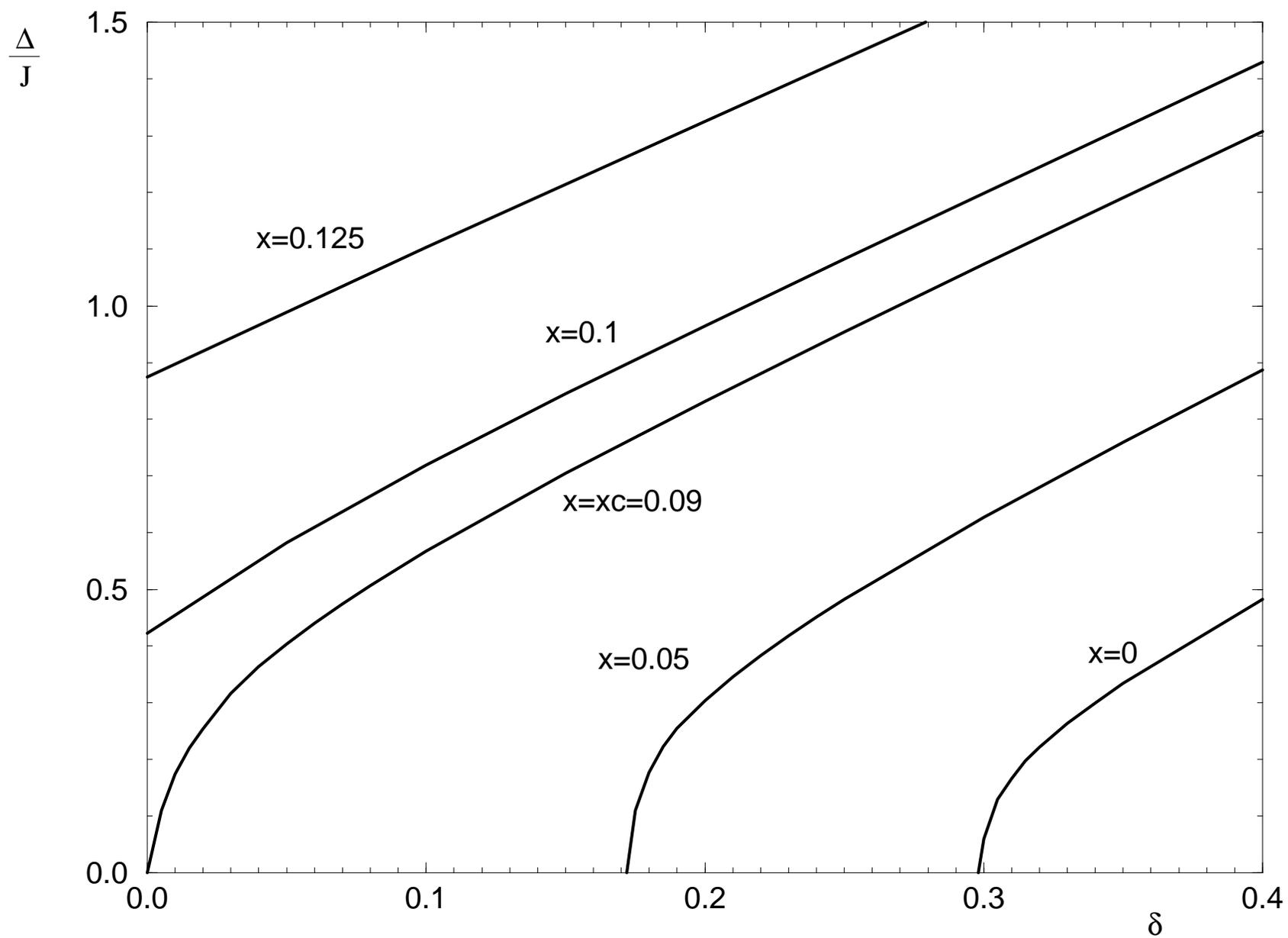
\begin{figure}
\caption
{The magnon "gap" (centre of gravity of spectral function) as a function of 
explicit dimerization $\delta$
for $t/J=3$ and different values of hole concentration $x$,
$\langle n_i\rangle =1-x$.}
\label{fig.2}
\end{figure}
 
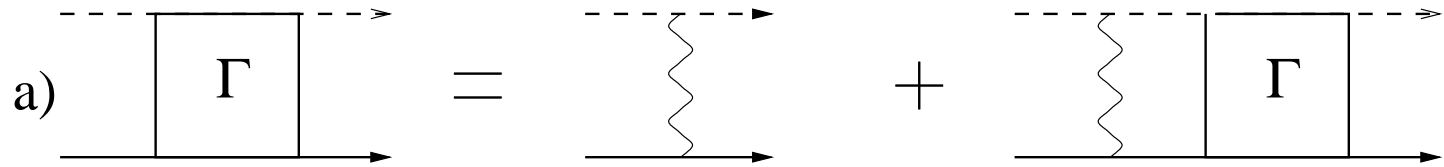
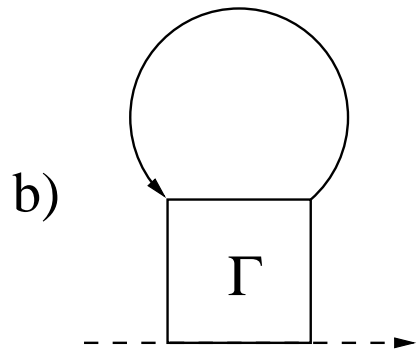
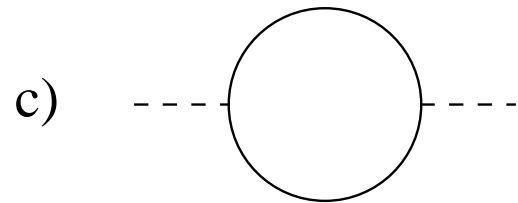
\begin{figure}
\caption
{(a) Bethe-Salpeter equation for hole-magnon scattering vertex $\Gamma$.
Solid line corresponds to the hole and dashed line to the magnon.
(b) Brueckner contribution to the magnon self energy.
(c) "triple" contribution to the magnon self-energy.}
\label{fig.3}
\end{figure}

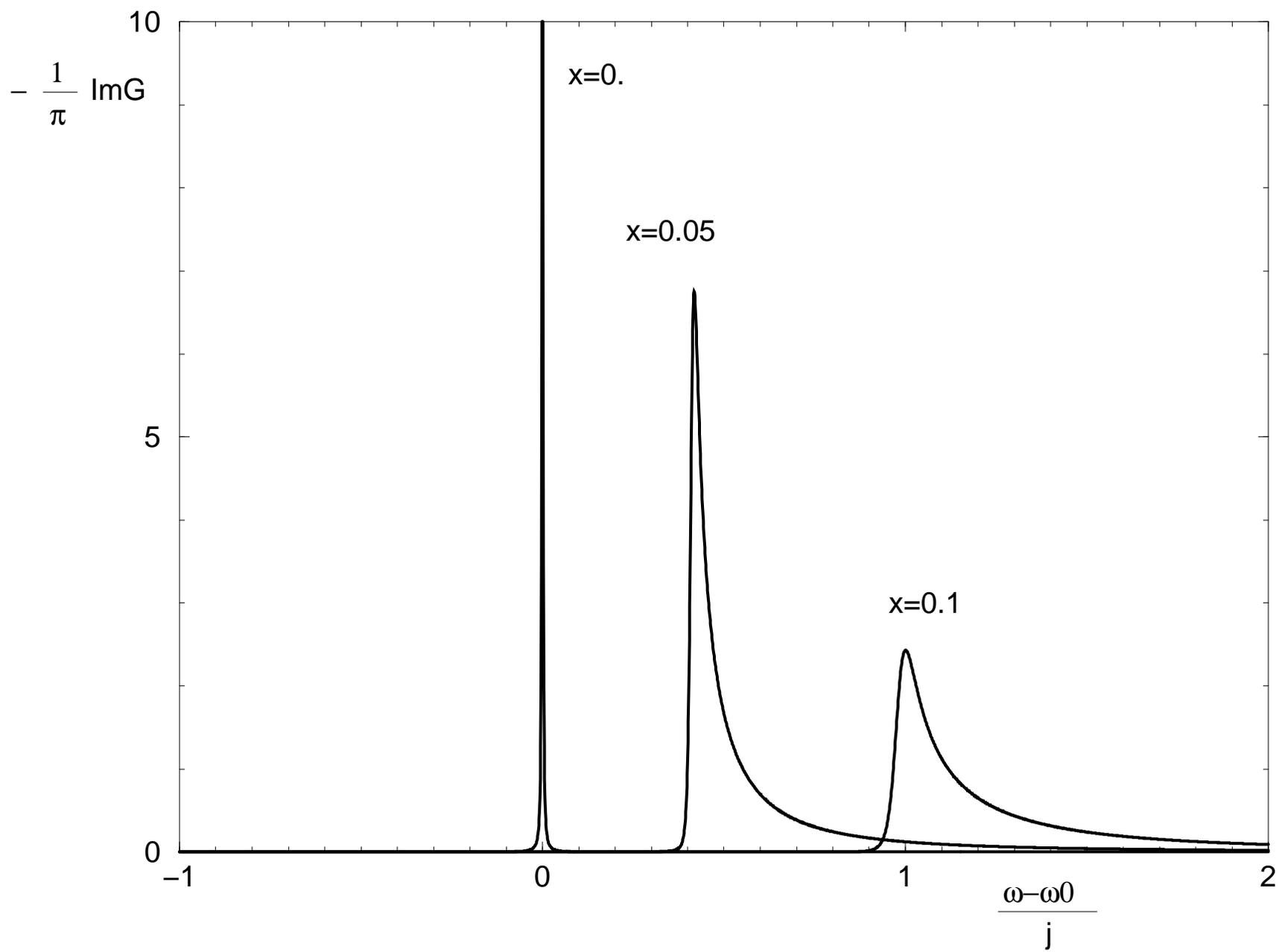
\begin{figure}
\caption
{Magnon spectral density at ${\bf k}={\bf k_0}=(0,\pi)$ in the limit 
$J_{\perp}\gg j$, $t/j=3$, and different hole concentrations.}
\label{fig.4}
\end{figure}

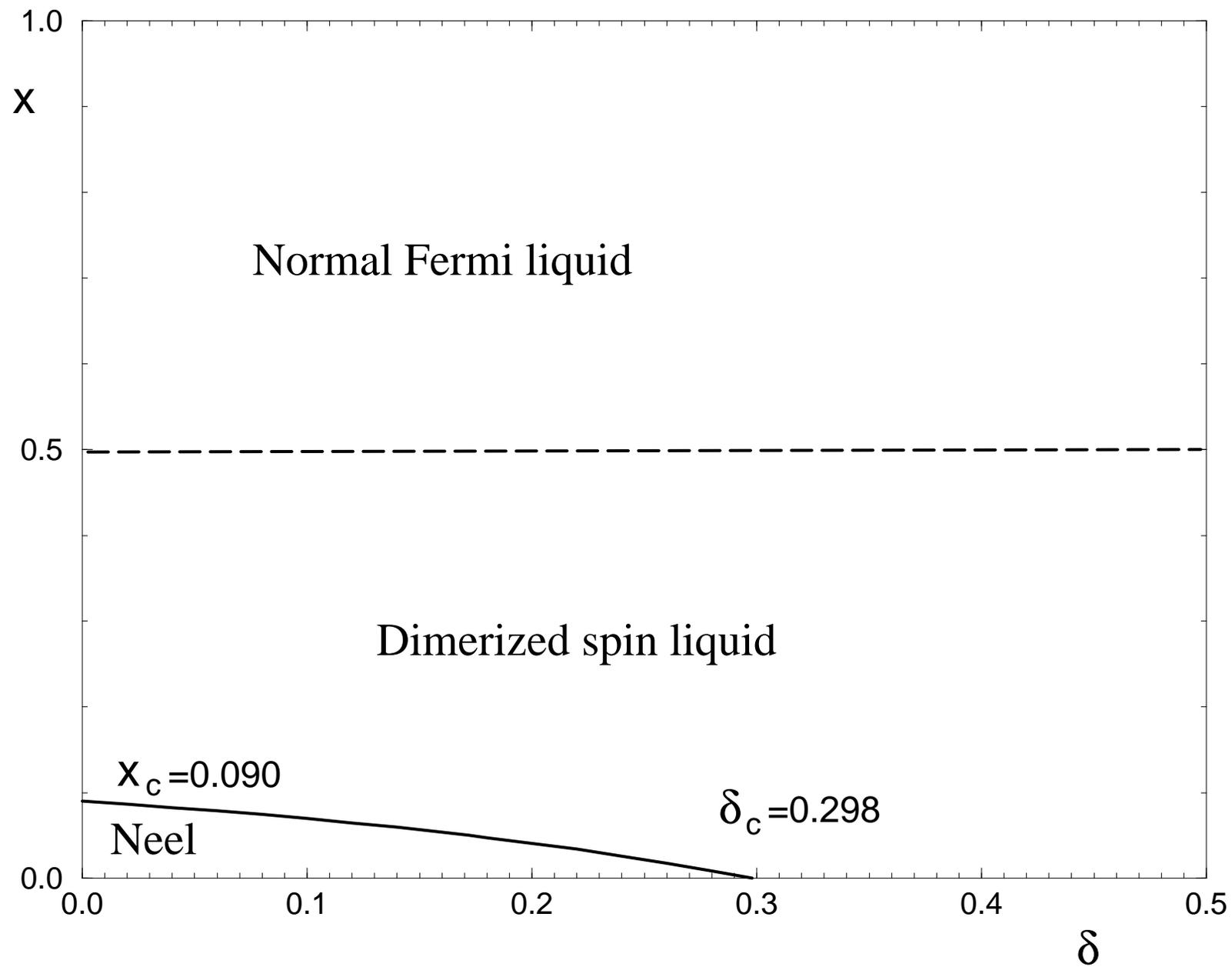
\begin{figure}
\caption
{Phase diagram of the $t-J-\delta$ model ($t/J=3$) in the plane doping ($x$) -
explicit dimerization ($\delta$).}
\label{fig.5}
\end{figure}


\begin{references}
\bibitem{com1} This interaction is responsible for
renormalization of the bare hole dispersion (\ref{ee}). Even for the
spin ladder, where the gas parameter $n_t \approx 0.3$,  this renormalization
is not large \cite {sus}. In the present case $n_t \lesssim 0.07$.
This is why the renormalization is very small.
\end{references}
\end{document}